\documentclass[%
 reprint,  amsmath,amssymb,
 aps,prl
]{revtex4-1}

\usepackage{graphicx}
\usepackage{dcolumn}
\usepackage{bm}
\usepackage{systeme,braket,color}
\usepackage{esint}
\usepackage[colorlinks=true,linkcolor=blue]{hyperref}

\newcommand{\addDS}[1]{\textcolor{black}{#1}}
\newcommand{\todo}[1]{\textcolor{black}{#1}}

\begin{document}

\preprint{APS/123-QED}

\title{Thresholdless excitation of edge plasmons by transverse current}

\author{Aleksandr S. Petrov}
\email{aleksandr.petrov@phystech.edu}
\author{Dmitry Svintsov}%
\affiliation{%
 Laboratory of 2D Materials' Optoelectronics, Moscow Institute of Physics and Technology, Dolgoprudny 141700, Russia
}%

\date{April 20, 2020}

\begin{abstract}

We theoretically demonstrate that dc electron flow across the junction of two-dimensional electron systems leads to excitation of edge magnetoplasmons. The threshold current for such plasmon excitation does not depend on contact effects and approaches zero for ballistic electron systems, which makes a strong distinction from the well-known Dyakonov-Shur and Cerenkov-type instabilities. We estimate the competing plasmon energy gain from dc current and loss due to electron scattering. We show that plasmon self excitation is feasible in GaAs-based heterostructures at $T\lesssim 200$ K and magnetic fields $B \lesssim 10$ T.

\end{abstract}

\maketitle

Edge plasmon is a collective electronic excitation propagating along the boundary of a two-dimensional electron system (2DES). Since their first observation~\cite{mast1985observation,glattli1985dynamical}, edge plasmons proved to be challenging yet fruitful phenomena to explore. The edge plasmons differ dramatically from their higher-dimensional counterparts: the former usually have longer lifetimes~\cite{MFP_of_EMPs,Peters_AF_plasmons}, manifest extraordinary light confinement~\cite{fei2015edge,andreev2017azbel} and exhibit unique chiral properties~\cite{mahoney2017zero,jin2017infrared,song2016chiral,Muravev_Crystal_for_EMPs} {\addDS{ such as unidirectional propagation}}. 
These features make edge plasmons promising information carriers in future {\addDS{ integrated circuits}}, 
but the technological progress is hindered by their laborious excitation. Thus, optical excitation techniques involve ponderous near-field equipment~\cite{fei2015edge} or additional sample processing (e.g., waveguide deployment~\cite{MFP_of_EMPs}), whereas electrical excitation of edge plasmons requires ultra-short pulses~\cite{ernst1996acoustic}.

In this Letter, we suggest a simple method for electrical excitation of edge plasmons in continuous regime: excitation by direct transverse current. This method complements the family of 
\addDS{current-driven} plasmon instabilities \addDS{in semiconductor heterostructures} containing Cerenkov-type ~\cite{krasheninnikov1980instabilities,mikhailov1998plasma}, beam~\cite{Kempa_beam,Gruzinskis_beam_instability} and Dyakonov-Shur~\cite{dyakonov1993shallow} instabilities. However, all mentioned cases concern the excitation of 2d plasmons by current co-propagating with excited wave. This resulted either in large threshold velocities for instability onset~\cite{mikhailov1998plasma}, or in extreme sensitivity to contact effects~\cite{crowne1997contact}. Accordingly, though current-driven electromagnetic emission in solids has been observed~\cite{kopylov_turbulence,el2010algan,Tsui_emission}, its relation to any plasmon instability is still debated~\cite{dyakonov2008boundary,Chaplik_abs_and_em,mendl2019coherent}.

\begin{figure}
    \centering
    \includegraphics[width=\linewidth]{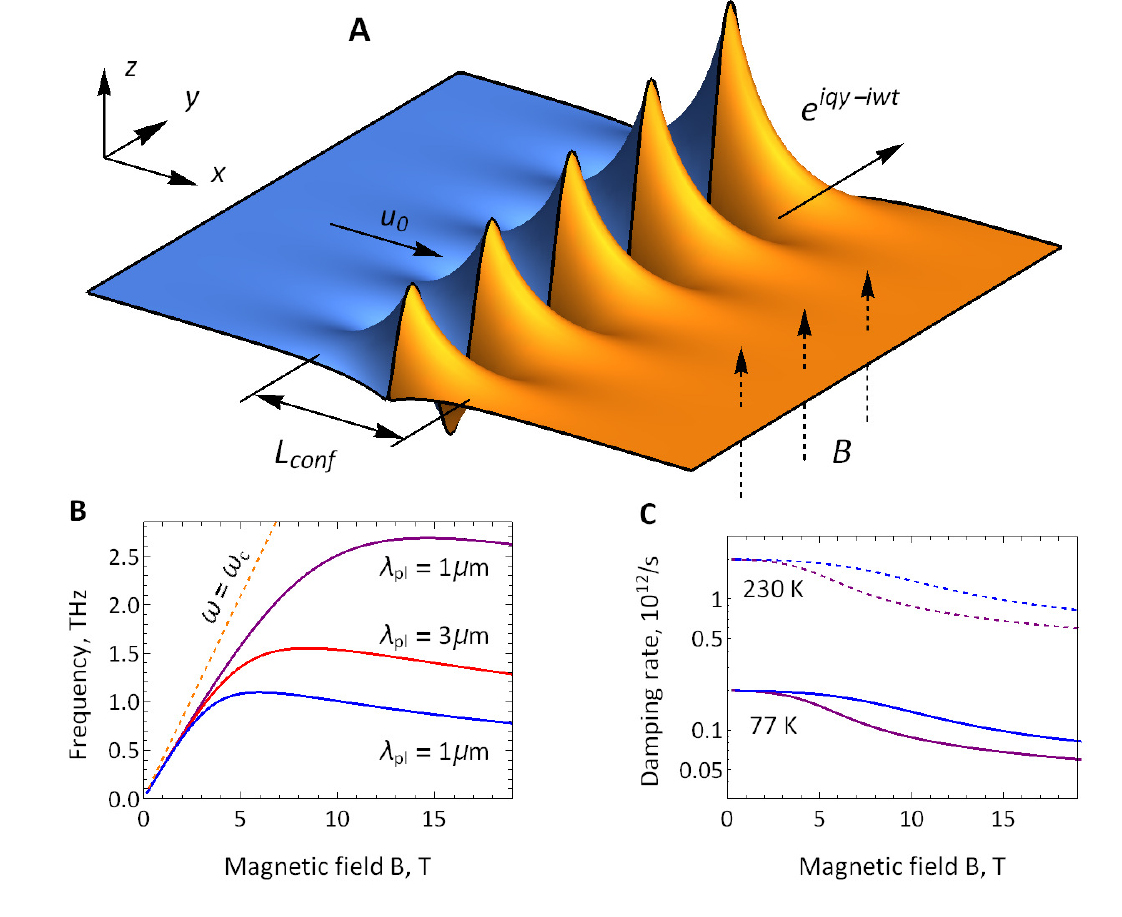}
    \caption{(A) Schematic of electric potential distribution for an inter-edge magnetoplasmon confined  between two conductive half-planes with characteristic confinement length $L_{\rm conf}$. The plasmon is chiral and propagates with wave vector $q>0$ if $n_r > n_l$ in magnetic field $B_z > 0$. The growing wave amplitude illustrates the gain from dc current $\mathbf{u}_0$; (B) IEMP spectrum for a $\rm GaAs/AlGaAs$ heterostructure ($m=0.067 m_e$, dielectric permittivity $\varepsilon = 1$ for simplicity) at different plasmon wave lengths $\lambda_{\rm pl} = 2\pi/q \simeq L_{\rm conf}/2$ . Carrier densities are $n_l=10^{11}\,\mathrm{cm}^{-2}$, $n_r=9\cdot10^{11}\,\mathrm{cm}^{-2}$. Orange dashed line stands for cyclotron frequency; (C) Damping rate dependence from magnetic field for IEMPs from panel B at different temperatures; effective momentum relaxation times were assumed to be 0.5\,ps for 230\,K and 5\,ps for 77\,K~\cite{schlom2010upward,andreev2014measurement}. Line colors correspond to plasmon wavelengths defined in panel B }
    \label{fig-IEMP-scheme}
\end{figure}

\addDS{The proposed technique for edge plasmon excitation has no current threshold for sufficiently clean systems and is insensitive to contact effects. It is inherited from a proposal of boundary instability in 2DES with fully imaginary (turbulent) spectrum~\cite{dyakonov2008boundary,petrov2016plasma}.} 
\todo{In this Letter, we show that turbulent plasma instability is the limiting case of a more general phenomenon -- instability of edge magnetoplasmons with properly defined spectrum.}
We develop a theory of current-driven edge plasmon instabilities, determine their frequencies and growth rates, and suggest a route for their experimental observation.


As an illustrative \addDS{and exactly solvable model}, we study \addDS{the effect of transverse electric current} on inter-edge magnetoplasmons (IEMP). These waves exist on the boundary between two conductive half-planes in an external magnetic field $B$~\cite{, mikhailov1992inter,kirichek1995magnetoplasmons,sommerfeld1995magnetoplasmons,sukhodub2004observation}. In what follows, we model the boundary as a step-like profile of electron density $n_0(x) = n_l\theta(-x) + n_r\theta(x)$. IEMP is a chiral mode with direction of propagation depending on direction of $B$ and carrier density contrast $n_r-n_l$. For definiteness, we choose $B>0$, $n_r>n_l$; in this case, the plasmon travels co-directional with 
the $y$-axis in Fig.~\ref{fig-IEMP-scheme}A. 

The spectrum of IEMPs is non-trivial: in weak fields its frequency is proportional to the magnetic field, while in strong fields the frequency acquires $\ln (\omega_c/\omega_{2d})/\omega_c$ dependence (Fig.~\ref{fig-IEMP-scheme}B), where $\omega_c$ is the cyclotron frequency and $\omega_{2d}$ is the plasma frequency of \addDS{unbounded 2DES}.
The dependence of wave damping on carrier momentum relaxation time $\tau_p$ is also noteworthy: in weak fields, the IEMP damping rate is 2 times higher than the usual $1/2\tau_p$ estimate for 2d and 3d plasmons, while in strong fields the damping rate is much lower and scales as $1/B$ (Fig.~\ref{fig-IEMP-scheme}C).

In what follows, we demonstrate that IEMPs \addDS{can be excited by the transverse electric current, and establish the general features of such an instability}. 
In our analysis we adopt the hydrodynamic model for electron transport~\footnote{Hydrodynamics not only provides the simplest framework for description of plasmons~\cite{fetter1985edge}. This model can be strictly derived from kinetic equation if wave frequency is well below the carrier-carrier collision frequency~\cite{Crossover,BGK-model}. Electron-phonon and electron-impurity collisions can also result in strong relaxation of non-hydrodynamic harmonics of distribution function, thereby effectively leading to hydrodynamic transport~\cite{Alekseev_NMR}.}.
In linearized form \addDS{with respect to variations of carrier density $n$ and drift velocity $\mathbf{u}$}, the hydrodynamic equations read
\begin{gather}
\label{eq-hd-full1}
    \partial_t  n + \nabla\left(n_0 \mathbf{u} + \mathbf{u}_0 n\right) = 0; \\
    \label{eq-hd-full2}
    \partial_t \mathbf{u} + \delta\left\{(\mathbf{U},\nabla)\mathbf{U}\right\} = -\frac{e}{mc}[\mathbf{u},\mathbf{B}] - \frac{e\mathbf{E}}{m}, 
\end{gather}
where $e>0$ is the elementary charge, $m$ is carrier effective mass, $c$ is the speed of light, $u_0 = u_l\theta(-x) + u_r\theta(x)$ is transverse drift velocity~\footnote{Generally speaking, the drift velocity is affected by the magnetic field and should have non-zero $y$-component. In the main text we treat the case when the (remote) $y$-boundaries of the sample have already accumulated a compensatory charge such that the velocity is directed solely along the $x$-axis.}, $\mathbf{E}=-\nabla\varphi$ is plasmon electric field, $\delta\left\{(\mathbf{U},\nabla)\mathbf{U}\right\} = (\mathbf{u}_0,\nabla)\mathbf{u} + (\mathbf{u},\nabla)\mathbf{u}_0$, square brackets denote vector product. \addDS{To find the eigen frequencies of plasmons, one supplements these equations with self-consistent field relation $\varphi(x) = -e \mathcal G[n] \equiv -e \int {\mathrm d\mathbf{r}' G(\mathbf{r},\mathbf{r}')n(\mathbf{r}')}$, where $G(\mathbf{r},\mathbf{r}')$ is the Green's function of Poisson's equation.}

The presence of carrier drift makes the conductivity tensor non-local in each of the half-planes, which significantly tangles the solution of the resulting eigenvalue problem (see, for example~\cite{cohen2018hall, margetis2020nonretarded}). Fortunately, analytical treatment is greatly simplified if we consider carrier drift as a small perturbation over the IEMP profile in an unbiased 2DES. This is done in the framework of a recently developed perturbation theory for hydrodynamic plasmons~\cite{petrov2019perturbation}.

This theory states that if $\lambda$th plasmon mode with frequency $\omega_\lambda$ is subject to a small perturbation $\hat{V}$,
then the perturbation-induced correction to the frequency is given by
\begin{equation}
    \label{eq-pert}
    \delta\omega_\lambda = \frac{\langle\mathbf{\Phi}_\lambda | \hat{H}\hat{V}\mathbf{\Phi}_\lambda\rangle}{\langle\mathbf{\Phi}_\lambda | \hat{H}\mathbf{\Phi}_\lambda\rangle},
\end{equation}
where $\hat{H}$ is the \addDS{''Hamiltonian operator'' governing the net energy of the wave}
\begin{equation}
\label{eq-Hamiltonian}
   \hat{H} = 
   \begin{pmatrix}
    e^2/m\,\mathcal G[\cdot] & 0 & 0  \\
   0 &  n_0(x) & 0 \\
   0 &  0 & n_0(x) \\
   \end{pmatrix},
\end{equation} 
\addDS{effect of current is described by the perturbation operator}
\begin{equation}
\label{eq-def-Vdrift}
    \hat{V} = -i\begin{pmatrix}
    					\partial_x[u_0(x)\cdot] & 0 & 0 \\
                        0 & \partial_x[u_0(x)\cdot] & 0\\
                        0 & 0 & u_0(x)\partial_x\cdot
    					\end{pmatrix},
\end{equation}
$\mathbf\Phi_\lambda = \{ n(x), u_x(x), u_y(x)\}^{\rm T} e^{iqy}$ is a three-dimensional vector comprising unperturbed plasmon charge density and velocity, and the inner product is defined as 
\begin{equation*}
\langle \Phi_\lambda |\hat{H} \Phi_{\lambda'} \rangle = \int d\mathbf{r}\left[ \frac{e^2}{m}n_\lambda^* \mathcal{G}[n_{\lambda'}] +n_0\mathbf{u}_\lambda^* \mathbf{u}_{\lambda'}\right].
\end{equation*}

\begin{figure}
    \centering
    \includegraphics[width=\linewidth]{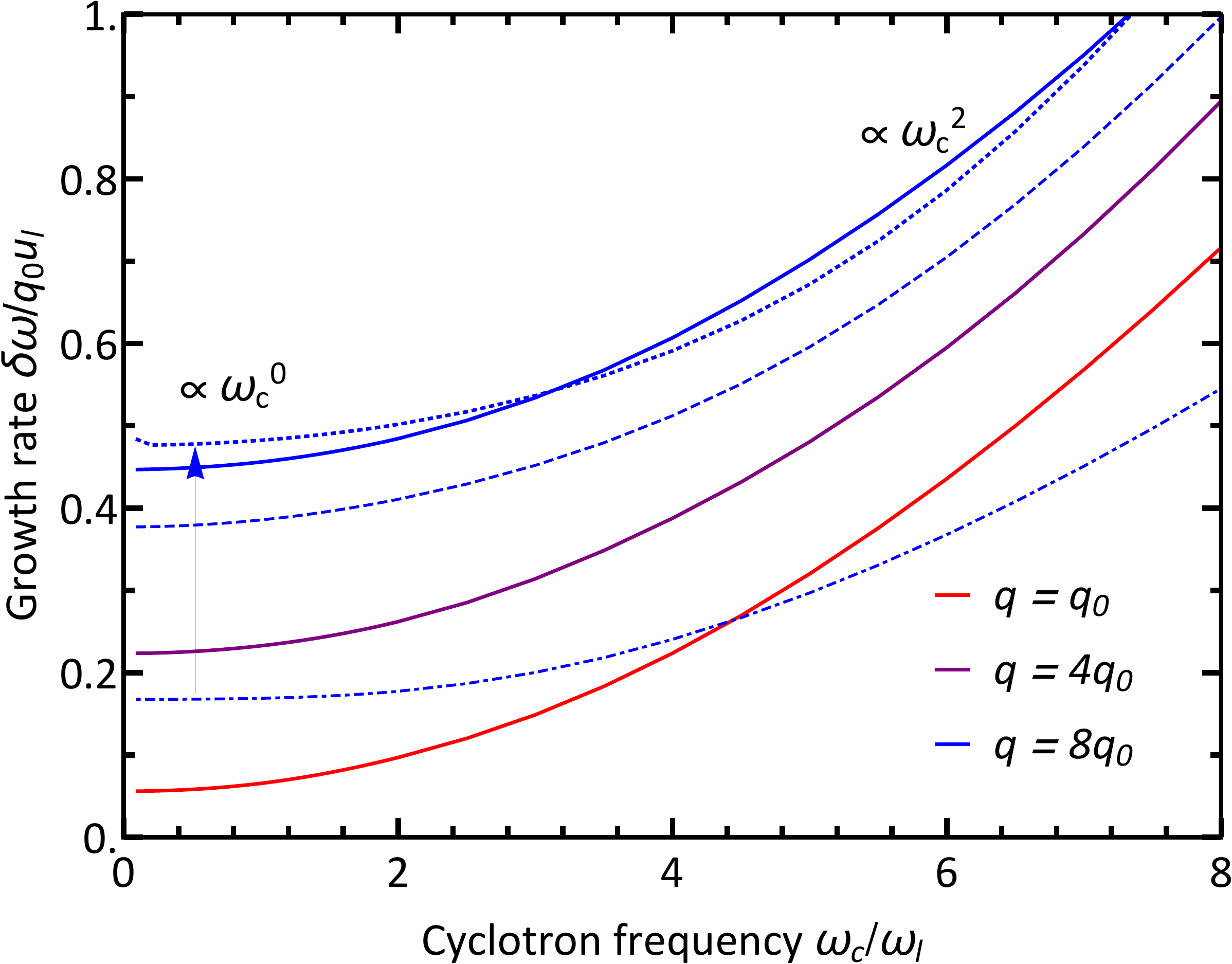}
    \caption{Calculated IEMP growth rate (in units of $q_0u_l$) vs cyclotron frequency (normalized by plasma frequency in the left half-plane $\omega_l(q_0)$) at various wave vectors and density contrasts for GaAs/AlGaAs heterostructure, $q_0 = 2\pi/(0.5\,\mu m)$, $u_l = 10^7\,$cm/s. Solid lines correspond to relative density contrast $n_r-n_l/n_r+n_l$ value $0.8$, dot-dashed -- $0.2$, dashed -- $0.6$, dotted -- $0.9$, while $n_l$ is fixed at $10^{11}\,\mathrm{cm}^{-2}$. The growth rate in weak fields saturates as the relative density contrast approaches 1 (blue arrow).}
    \label{fig-growth rate}
\end{figure}


We \addDS{managed to evaluate the current-induced perturbation (\ref{eq-pert}) of plasmon frequency in symbolic form for IEMPs at the step-like discontinuity in carrier density. This procedure results in}
\begin{equation}
    \label{eq-corr-general}
    \delta\omega_{\rm emp} = -ij_0\frac{\left.\left[\dfrac{m |u_x|^2}{2} - \dfrac{e^2E_x^2}{2m\omega^2} \right]\right|_{-0}^{+0}}{\int\limits_{-\infty}^\infty  [mn_0\left(|u_x|^2 + |u_y|^2\right) - e\varphi n ]dx},
\end{equation}
\addDS{where the notation $[...]|_{-0}^{+0}$ stands for discontinuity of the quantity across the interface,} and $j_0 =n_lu_l = n_ru_r$ is carrier flux.



The correction to plasmon frequency (\ref{eq-corr-general}) is purely imaginary, which corresponds to wave self-excitation for $\mathrm{Im}\,\delta\omega_\lambda>0$, and damping for $\mathrm{Im}\,\delta\omega_\lambda<0$. It depends linearly on current $j_0$ which is a natural consequence of perturbation theory.  From the above equation we readily reveal the necessary conditions for edge plasmon excitation by direct current. First, plasmons cannot be excited in the absence of magnetic field; the latter tangles $u_x$ velocity component with perpendicular electric field $E_y$ leading to non-zero numerator. Highly symmetrical modes are insensitive to drift as well. The example of such a mode is proximity plasmon bound between homogeneous 2DES and metallic electrode~\cite{muravev2019two,zabolotnykh2019interaction}. 


To \addDS{judge on} the definite effect of drift, we plug the known \addDS{distributions of fields 
in the IEMP mode}~\cite{mikhailov1992inter} into Eq.~(\ref{eq-pert}) and numerically evaluate the integrals (see SI for the procedure). As a result, we obtain the IEMP growth rate dependence on the cyclotron frequency shown in Fig.~\ref{fig-growth rate}.

\begin{figure}
    \centering
    \includegraphics[width=\linewidth]{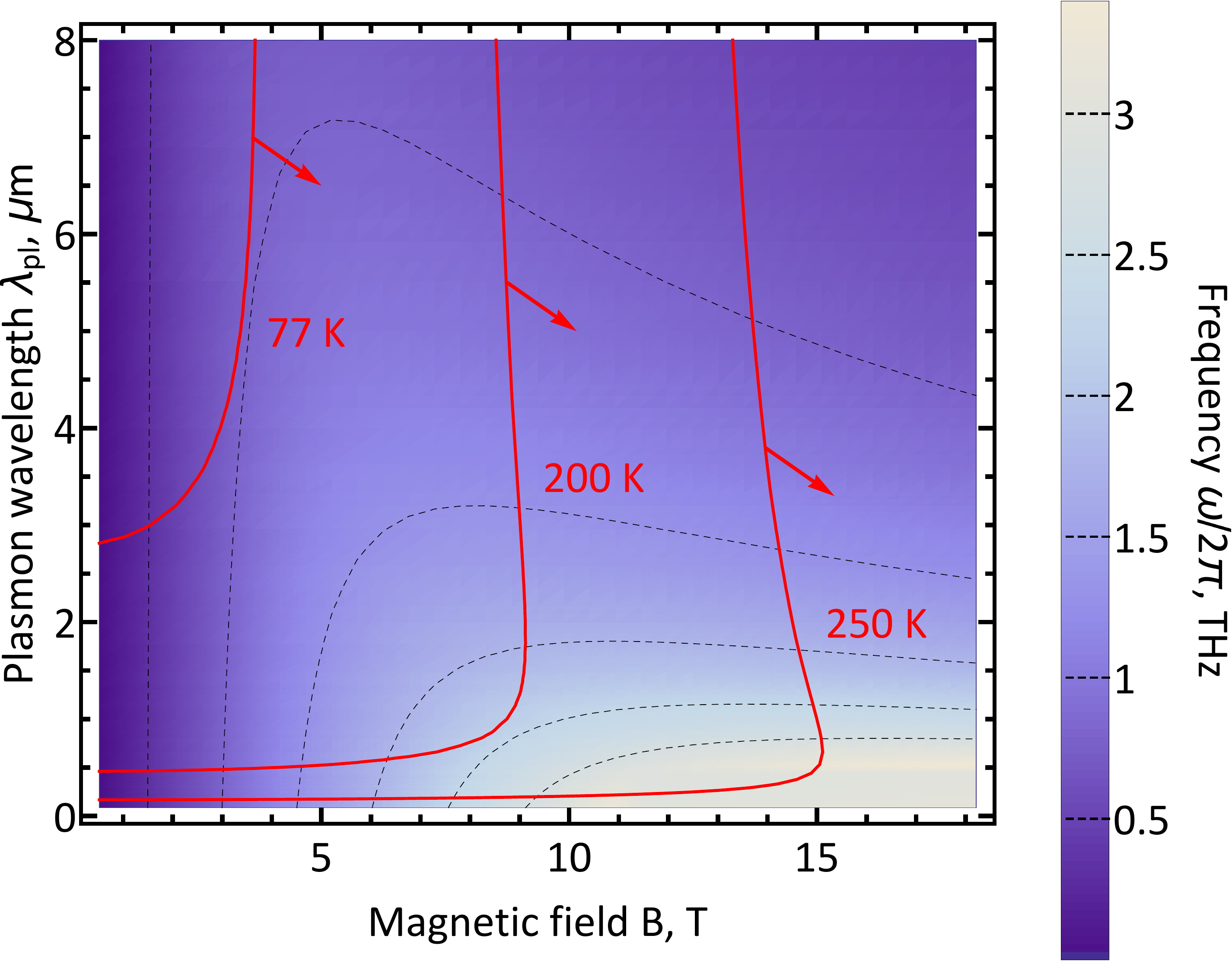}
    \caption{\addDS{Color map of edge magnetoplasmon dispersion $\omega_{\rm emp}(\lambda_{\rm pl},B)$ overlaid with ''critical lines'' of instability calculated at three different temperatures for $\rm GaAs/AlGaAs$ heterostructure. Waves with parameters to the right from ''critical lines'' have threshold carrier velocity below the saturation velocity in GaAs ($\sim2\cdot10^7$\,cm/s). }Structural parameters are the same as in Fig.~\ref{fig-IEMP-scheme}B, effective momentum scattering time are the following: 5\,ps for 77\,K, 0.75\,ps for 200\,K and 0.25\,ps for 250\,K. }
    \label{fig-threshold}
\end{figure}

We observe that the instability benefits from pronounced density contrast at the boundary (see the blue arrow on Fig.~\ref{fig-growth rate}), and its behavior drastically differs in limits of weak and strong magnetic fields. In weak fields the instability growth rate scales linearly with the wave vector and is independent of $B$. In strong fields, the plasmon growth rate scales as $B^2$ and is independent of the wave vector. The growth rates in these limiting cases are given by:
\begin{gather}
    \label{eq-correction-weak}
    \delta\omega_{w} \simeq iqj_0\frac{n_r-n_l}{2n_rn_l}\propto \omega_c^0q^1\Delta n^1;\\
    \label{eq-correction-strong}
    \delta\omega_{s} \simeq -iqj_0\frac{\omega_c}{\omega}\frac{\left(n_r-n_l\right)^2\left(n_r+n_l\right)}{8n_r^2n_l^2}\propto \omega_c^2q^0\Delta n^1,
\end{gather}
where $\omega_c = eB/mc$ is the cyclotron frequency.

The instability \addDS{has zero threshold current in clean 2DES. In realistic systems}, it is mainly hampered by carrier scattering on phonons or impurities. 
Thus, it is important to estimate the threshold drift velocity $u_{th}$ at which gain (\ref{eq-corr-general}) takes over scattering loss.

To provide a quantitative picture, we examine the stability of dc current in $\rm GaAs/AlGaAs$ heterostructure for a wide range of magnetic fields and wavelengths. In Fig.~\ref{fig-threshold}, we plot the boundaries separating stability and instability regions at three temperatures, the instability regions are indicated by red arrows. The boundary lines are calculated from the balance between damping rate at a given wavelength and magnetic field, and the growth rate at GaAs saturation velocity ($\sim 2\cdot 10^7$cm/s) \footnote{Strictly speaking, Fig.~\ref{fig-threshold} does not take into account the mutual drift-damping influence: the growth rate was calculated on the basis of collisionless unperturbed state ($\tau_p = \infty$), while the damping rate was taken from unbiased state ($u_l=u_r=0$, see Fig.~\ref{fig-IEMP-scheme}C). This may affect the regions where momentum loss is most crucial: $\mathrm{Re}\,\omega\ll\mathrm{Im}\,\omega$, in Fig.~\ref{fig-threshold} it corresponds to $B\lesssim 1$\,T. }. We observe that IEMP can be easily excited at $77\,$K; its excitation at higher temperatures is possible for shorter wavelengths and/or stronger magnetic fields. However, it is not the absolute value of the magnetic field that governs the instability growth rate; instead, it is the $\omega_c/\omega_{2d}$ ratio. Hence, in order to achieve pronounced growth rates one can not only increase the field, but also decrease the fundamental 2DEG frequency (e.g., by depletion of carrier density). For example, electron gas on a surface of liquid helium usually exhibits $\omega_c/\omega_{2d}\simeq 1000$ even at $B=1$\,T~\cite{glattli1985dynamical}, which enormously boosts the quadratically-scaled growth rate (\ref{eq-correction-strong}).

We stress that edge plasmon instability should be distinguished from the Dyakonov-Shur instability. \addDS{The latter relies on the surplus of energy gained by plasmon at source over the energy lost at the drain, thus} being extremely sensitive to boundary conditions~\cite{cheremisin1999d}. In contrast, edge plasmon instability is independent of contact effects, as \addDS{the required energy transfer from dc current to plasmon occurs in the interior of 2DES in the vicinity of the density step.} What is more, the frequency of the excited plasmon is independent of sample length or width provided they significantly exceed plasmon wavelength. These features make IEMP instability a prominent candidate for creation of resonant-tunable arrays of plasmonic THz emitters.

It is remarkable that current-induced frequency shift can be obtained purely from energy conservation considerations (see Appendix), similarly to the Reynolds-Orr energy equation known in the fluid turbulence theory~\cite{reynolds1895iv,orr1907stability}. However, the strong inhomogenity of dc current flow ($\partial u_{0x}/\partial y\neq  0$) necessary for turbulence onset in fluids is not required for edge plasmon instability due to non-zero compressibility of electron system. 


Substantially, edge plasmon instability is just one of numerous manifestations of the flux-to-perturbation energy transfer 
in plasmonics. For example, it can be used to excite chiral plasmons without magnetic field~\cite{song2016chiral}, inter-surface magnetoplasmons (3d analog of IEMPs)
, higher-order (quadrupole, etc.) magnetoplasmon modes bound to a smooth edge~\cite{aleiner1994novel}, or increase the lifetime of decaying modes such as the upper mode of IEMP \cite{mikhailov1992inter}. It would be of particular interest to examine the stability of proximity magnetoplasmons \cite{muravev2019two,zabolotnykh2019interaction} with respect to external source drain-bias due to relatively simple experimental setup (no need for density contrast). Essentially, the magnetic field will be needed to break the proximity mode symmetry and make it susceptible to drift. 


In conclusion, we predicted thresholdless current-driven edge plasmon instability. Possible applications include electrical excitation of edge plasmons in continuous regime and creation of competitive resonant THz sources. The underlying mechanism for the reported instability is flow-to-perturbation energy transfer that proves to be a general phenomenon in plasmonics and has many potential manifestations.

\section*{Acknowledgement}
The authors are grateful to V. Muravev, M. Dyakonov, I. Zagorodnev and G. Alymov for fruitful discussions and comments. The authors acknowledge support from Russian Foundation for Basic Research, project No. 18-37-00206, and Foundation for the Advancement of Theoretical Physics and Mathematics “BASIS”, project No. 18-1-5-66-1 (development of perturbation theory for edge plasmons). Analysis of instability threshold was supported by the grant 18-72-00234 of the Russian Science Foundation.  

\appendix

\section{Flow-to-perturbation energy transfer}


We multiply the Euler equation (\ref{eq-hd-full2}) with $n_0\mathbf{u}$, integrate over the whole 2DES and eliminate the boundary contributions by Gauss-Ostrogradsky formula that can be done exceptionally by virtue of the localized nature of edge plasmon. Thus, we obtain the following equation for energy balance:
\begin{multline}
    \label{eq-energy-balance}
    \partial_t \int\left(K+\Pi\right)dS = \\ =  \int \frac{e(\mathbf{E},\mathbf{u}_0)}{m} n - n_0\mathbf{u}\,\delta\left\{(\mathbf{U},\nabla)\mathbf{U}\right\}  dS,
\end{multline}
where $$K+\Pi = \frac{n_0 \mathbf{u}^2}{2} - \frac{en \varphi}{2m}$$ is total plasmon energy density. Hence, plasmon energy density changes in time due to its interaction with stationary flow (right-hand side). We stress that electron compressibility is crucial for plasmon excitation; otherwise, the right-hand-side of Eq.~(\ref{eq-energy-balance}) vanishes for usual flows ($\partial_i v_{0j} = 0, i\ne j$). Remarkably, the perturbation theory result (\ref{eq-corr-general}) can be obtained by time-averaging of the energy balance equation (\ref{eq-energy-balance}) and expanding it to the first power of drift velocity.

\bibliography{refs}

\newpage

\begin{widetext}
\section{Calculation of drift-induced correction to IEMP spectrum}
Expansion of matrix elements in Eq.~(\ref{eq-pert}) leads to:
\begin{equation}
    \label{eq-pert full}
    \delta\omega = iqj_0\frac{-\omega_c^2/2\left(E(+0)^2-E(-0)^2\right) + \omega\omega_c\left(E(+0)-E(-0)\right)\varphi(0)}{2\omega^2 \fint\limits_{-\infty}^{\infty}n_0(t)(\varphi'(t)^2 + \varphi(t)^2)\,dt + (3\omega\omega_c-\omega_c^3/\omega)(n_l-n_r)\varphi(0)^2},
\end{equation}
where $t = q x$ is dimensionless coordinate, $\fint = \int\limits_{-\infty}^{-0} + \int\limits_{+0}^{+\infty}$,
\begin{gather}
    E(\pm 0) = \pm\frac{\omega_r^2 - \omega_l^2}{2\omega_{r,l}^2}\frac{\omega_c\pm\omega}{\omega},
\end{gather}
$\varphi(t)$ is IEMP profile~\cite{volkov1988edge}, and prime denotes derivative. The profile $\varphi(t) = \varphi_l\theta(-t) + \varphi_r\theta(t)$ can be reduced to:
\begin{equation}
    \varphi_{r,l}(t) = -\frac{\varphi(0)}{\pi}\frac{1-\omega_c/\omega}{X(i)}\frac{\omega_r^2-\omega_l^2}{\omega_c^2-\omega^2}\int\limits_1^\infty\frac{e^{\mp\xi t}\, d\xi}{\sqrt{\xi^2-1}}\frac{X(\mp i\xi)}{1+(\xi^2-1)/\alpha_\pm^2},
\end{equation}
where $\alpha_\pm = (\omega_c^2-\omega^2)/\omega_{r,l}^2$, 
\begin{equation}
    X_\pm(\xi) = \exp\left(-\frac{1}{2\pi i}\int\limits_{-\infty}^\infty\frac{d\xi'}{\xi'-\xi\mp i0}\ln\frac{\varepsilon_r(\xi')}{\varepsilon_l(\xi')}\right),
\end{equation}
$\varepsilon_{r,l}$ are the dielectric permittivities of right and left half-planes.

The main obstacle in numerical evaluation of the correction (\ref{eq-pert full}) is calculation of the $\fint$ integral. Luckily, analytical treatment is possible if we approximate the smooth function $X(\mp i\xi)$ by its value at the point $\xi=1$, where the integrand has a singularity. Then, after some simplifications we arrive at:
\begin{equation}
    \label{eq-profile-simplified}
    \varphi_\pm(t) = \frac{\varphi(0)}{\pi}\frac{\omega_r^2-\omega_l^2}{\omega(\omega_c\pm\omega)}\int\limits_1^\infty\frac{e^{\mp\xi t}\,\mathrm d\xi}{\sqrt{\xi^2-1}}\frac{1}{1+(\xi^2-1)/\alpha_\pm^2}.
\end{equation}

In order to evaluate the $\fint$ integral with the profiles (\ref{eq-profile-simplified}), we represent them as triple integrals (the prefactor is omitted):
\begin{gather}
    \label{eq-triple-integral}
    \int_0^\infty \varphi_+^2\, dt = \int_0^\infty dt\int_1^\infty d\xi_1\int_1^\infty d\xi_2\; \phi_+(\xi_1,t)\phi_+(\xi_2,t),
\end{gather}
and an analogous expression for the integral of $\varphi_+'^2$; $\phi_\pm(\xi,t)$ denote the integrands in Eq.~(\ref{eq-profile-simplified}). The integration over $dt$ is readily done -- it is just the integral from exponent product. The integrals over $d\xi_1$ and $d\xi_2$ are taken analytically by Wolfram Mathematica, except for one term in both cases. We arrive at:

\begin{gather}
    \label{eq-phi int}
   \int_0^\infty \varphi_+^2\,dt = \frac{\pi\alpha_r}{4(\alpha_r^2-1)}-\frac{\pi\alpha_r^2\,\mathrm{arcsh}(\sqrt{\alpha_r^2-1})}{4(\alpha_r^2-1)^{3/2}} + \alpha_r^4\int_1^\infty\frac{d\xi_2}{\xi_2^2-1}\frac{\mathrm{arcch}(\xi_2)}{(\alpha_r^2+\xi_2^2-1)^2}; \\
   \label{eq-E int}
   \int_0^\infty {\varphi'}_+^2\, dt = \frac{\pi\sqrt{\alpha_r^2-1}}{8}\ln\left(-1+2\alpha_r(\alpha_r+\sqrt{\alpha_r^2-1})\right) + \frac{\pi}{4}\left(\alpha_r+\frac{\mathrm{arcsh}\left(\sqrt{\alpha_r^2-1}\right)}{\sqrt{\alpha_r^2-1}}\right) + \alpha_r^4\int\limits_1^\infty\frac{\xi_2^2\, d\xi_2}{\xi_2^2-1}\frac{-\mathrm{arcch}(\xi_2)}{(\alpha_r^2+\xi_2^2-1)^2}.
\end{gather}

The final answer for Eq.~\ref{eq-triple-integral} is the sum of Eqs. (\ref{eq-phi int}) and (\ref{eq-E int}); the sum of the remaining integrals is taken via residues. The resulting expression is cumbersome, however its expansions in weak (\ref{eq-correction-weak}) and strong (\ref{eq-correction-strong}) magnetic fields are neat, see Eqs.~(\ref{eq-correction-weak}) and (\ref{eq-correction-strong}).

\end{widetext}

\end{document}